\documentclass[preprint,preprintnumbers]{revtex4}

\usepackage{graphicx}

\usepackage{amssymb}

\usepackage{mathrsfs}

\begin{document}
\title{Proposal for the  detection  of  Majorana  Fermions in Topological Superconductors }

\author{D. Schmeltzer}

\affiliation{Physics Department, City College of the City University of New York,  
New York, New York 10031, USA}
  

\begin{abstract}

\noindent  
One of the goals of modern spectroscopy   is to invent  techniques      which  detect neutral excitations  that have been theoretically proposed.
For superconductors, two-point transport measurements detect the Andreev crossed reflection which  confirms the existence of Majorana fermions.
Similar information can also be obtained from a measurement using two  piezoelectric transducers. One  transducer measures  the stress tensor response from the strain  field  generated by the second transducer. The ratio between the stress response and strain velocity determines the dissipative response. We show that the  dissipative   stress response can be used for studying excitations in a topological superconductor. We investigate a topological superconductor  for the case when an  Abrikosov vortex lattice is formed.  In this case the  Majorana fermions  are dispersive, a fact that is used to compute  the  dissipative stress response.


\end{abstract}

\maketitle


\vspace{0.2 in}

\vspace{0.2 in}

\textbf{I. Introduction}

\vspace{0.2 in}

 The proximity of a superconductor \cite{FuKane} to the surface of a topological insulator (TI)  gives rise to a topological superconductor (TS)  characterized by  the Majorana zero modes.  Recently,  vortices and  Majorana fermions in a magnetic field  have been  reported  \cite{J-P.Xu} in heterostructures  of  Bi$_{2}$Te$_{3}$/NbSe$_{2}$.  Additionally,  Majorana fermions have been studied  in Abrikosov lattices  \cite{Stern,Franz}.
 The Majorana zero modes are neutral excitations which, in  an Abrikosov vortex lattice, become a  gapless dispersive band \cite{Biswas,Stern,Franz}. 
In a variety of materials a scanning tunneling microscope (STM) is used to detect charge tunneling. The spin tunneling information
is usually obtained from a  Magnetic Force Microscope  (MFM) \cite{Hoffman,Balatzky}.
Recently it was proposed that the Majorana fermions can be  detected by combining STM and AFM \cite{Loss,Yazdani,Science}.

The question which we pose here and needs to be answered is: Is it possible for a single  STM  measurement to detect the Majorana fermions?
 Two Majorana modes located  at the  two ends of a $p$-wave  (or equivalently a one-dimensional   wire with spin-orbit interaction in proximity of an $s$-wave superconductor and magnetic field) wire might be detectable  by two STM  experiments (two STM  tips separated in space). It is important to mention that   the  Andreev crossed reflection \cite{Flensberg,DavidMajorana}, in which an electron is incident on one side of the superconductor and a hole comes out at the other end, has been measured in a two-leads experiment. This suggests that two STMs  alone are able to detect Majorana fermions.
 
Here we do not discuss the two-tip experiment since it is similar to the Andreev reflection experiment, instead we consider the sound wave analog.
The  analogous  configuration for two STMs tips  can be  realized   by two piezoelectric transducers. One transducer is used to generate a {\it strain} wave, whereas the other is used to detect the {\it stress}  response. The stress response is equivalent to the electromagnetic (paramagnetic) response of a superconductor to an electric  field \cite{Schrieffer}. We show here that for a  TS the  stress response also probes the  Majorana modes. 

A number  of methods based on sound waves have  been used to investigate superconductors \cite{Maali,Maynard,Aref,Magnusson}.
 Ultrasound  attenuation studies  \cite{Kadanoff,Tsuneto},  and   investigations of   the  $p$-wave superconductor Sr$_{2}$RuO$_{2}$ have been carried out  in ref. \cite{Maki}.
In  the  late fifties  ultrasound attenuation techniques were used to measure the  temperature dependence of the  superconducting  gap   \cite{Tinkham,Schrieffer,Philip,Rodriguez,Tremblay} and recently  the techniques have been  applied to  liquid $^{3}$He \cite{Nagai}.

The purpose of this paper is to demonstrate that piezoelectric transducers can be used to  detect  Majorana fermions in an Abrikosov vortex lattice.  The tunneling amplitude  of the   Majorana  fermions   gives rise to a dispersive band \cite{Biswas} which is detectable. 
 We compute  the stress  {\it viscosity} as  a response to an applied   strain velocity field  \cite{LandauI,LandauII}.  The  explicit dependence of the strain field on the system is obtained from a coordinate transformation  \cite{Katanaev}. The  linear  stress  response  theory \cite{Fetter}  used for a TS provides  information   about the Majorana fermions.   
In order to demonstrate the  detection of Majorana fermions we consider  a TS  \cite{Ivanov,FuKane,Yakovenko,J-P.Xu,Stern,Franz} 
 which has an {Abrikosov vortex lattice}.  

 In this paper we have derived the following specific results: 
(a) We have obtained the vortex lattice solution for a $p$-wave superconductor.
(b) We have derived the stress-strain Hamiltonian  and have computed the stress viscosity using the linear response theory.
(c) We have identified the sound analog of the Andreev crossed reflection and have obtained the viscosity equivalent to the crossed reflection conductance.

The structure of the paper is as follows: In Sec. II we show that for an attractive interaction on the surface of a TI a TS is obtained. In the presence of an Abrikosov vortex lattice dispersive Majorana fermions are formed.
In Sec. III we present  the formation of the Abrikosov vortex lattice. We  find   dispersive  Majorana fermions  and quasi-particles  in the vortex lattice. A new solution for the $p$-wave  {Abrikosov  vortex lattice} is  obtained and discussed in detail.
Section IV  is devoted to the derivation of the viscosity tensor for a  TS.
In Sec.V  we compare our  results  to the one obtained  by the ultrasound attenuation technique.
In Sec.VI we compute  the transverse  impedance for  the TS  in a magnetic field.  The transverse impedance provides distinct  information about the Abrikosov   vortex lattice and  the  Majorana fermions. Section VII is devoted to comments  on the Andreev crossed reflection induced in one dimension  by a  piezoelectric transducer.
Section VIII  contains our main conclusions.

\vspace{0.2 in}

\textbf{II. Formation of  Majorana fermions  on the surface  of a  TI  with attractive interactions} 

\vspace{0.2 in}

In this section we review the formation of a $p$-wave superconductor on the surface of a TI with attractive interactions.
 For two space dimensions,  the quantum Hall system  and the $p$-wave superconductor \cite{Ivanov,FuKane,Yakovenko} are characterized by the first Chern integer  number $C^{1}$ (which means that the integral of the Berry curvature over a closed manifold is quantized in units of $2\pi$) \cite{Taylor}.
In the presence of an attractive interaction  (due to the electron-phonon interaction   or  proximity  to another superconductor), on the  surface of  a TI a two-dimensional  TS emerges.
 The proximity of a superconductor \cite{FuKane} to the  three dimensional TI  gives rise  to Majorana   zero modes  on the surface of the TI.  Recently  it was reported  that the application  of a magnetic field  on   the heterostructure Bi$_{2}$Te$_{3}$/NbSe$_{2}$ induces an Abrikosov vortex lattice \cite{J-P.Xu}.
 
We propose  that a  realization of  the  model  introduced in \cite{FuKane}  emerges    from the  TI surface   Hamiltonian $h(\vec{k})=-\sigma_{2}k_{1}+\sigma_{1}k_{2}$ in the presence of an attractive interaction. We express the pairing interaction in terms of the field $\psi_{\sigma}(\vec{x})= \sum_{k}e^{i\vec{k}\cdot \vec{x}} u^{(+)}(\vec{k})_{\sigma}$ where    $ u^{(+)}(\vec{k})_{\sigma}$ are the TI surface  spinors  for the conduction band,  $ u^{(+)}(\vec{k})=\frac{1}{\sqrt{2}}\Big[1,i\frac{k_{1}-ik_{2}}{|\vec{k}|}|\Big]^{T}\approx \frac{1}{\sqrt{2}}\Big[1,i\frac{k_{1}-ik_{2}}{k_{0}}\Big]^{T}$ ($T$ stands for transpose),  and $k_{0}$  is a momentum scale. This representation generates the linear derivatives of  the pairing field. For  a  positive chemical potential $\mu>0$   in the presence  of a  magnetic field, it gives rise to a superconductor with  vortices.
The attractive interaction expressed in terms of the TI spinors gives rise to the $p$-wave Hamiltonian \cite{FuKane} which in our case also includes vortices.
\begin{eqnarray}
&&H_{T.I.+SC.}= \int\,d^2x\Big[ C^{\dagger}(\vec{x})\frac{ v}{2k_{F}}\Big((-i\vec{\partial}_{x} -\frac{e}{\hbar}\vec{A}(\vec{x},t))^2- k^2_{F}\Big)C(\vec{x})+
\frac{\Delta^{*}(\vec{x},t)}{2k_{0}} C(\vec{x},t)\Big(\partial_{1}-i\partial_{2}\Big)C(\vec{x},t)  \nonumber\\ &&    - \frac{\Delta(\vec{x},t)}{2k_{0}}             C^{\dagger}(\vec{x},t)\Big(\partial_{1}+i\partial_{2}\Big)C^{\dagger}(\vec{x},t) +\frac{1}{g} |\Delta (\vec{x},t)|^2\Big]. \nonumber\\ &&
\end{eqnarray}
  The pairing field  $\Delta(\vec{x})$  depends  on the phase  $\theta(\vec{x})$ which   includes a multivalued part.  We perform the  gauge transformation:
$C(\vec{x})=e^{\frac{-i}{2}\theta(\vec{x})}\hat{C}(\vec{x})$,
 $C^{\dagger}(\vec{x})=
\hat{C}^{\dagger}(\vec{x})e^{\frac{i}{2}\theta(\vec{x})}$ and the Hamiltonian $H$ is replaced by $\hat{H}$.
The pairing field $\Delta(\vec{x})$ has points $\vec{x}\approx\vec{R}_{k}$, $k=1, 2, ...$, where $\Delta(\vec{x})$  vanishes, 
$\Delta(\vec{x})\approx |\Delta|e^{i\theta(\vec{x})}$, $\Delta^{\dagger}(\vec{x})=|\Delta|e^{-i\theta(\vec{x})}$, where  $\theta(\vec{x})\equiv\theta(\vec{x};\vec{R}_{0}=0,\vec{R}_{1},...,\vec{R}_{k},...)$  is the  multivalued phase.
As a result of the gauge transformation the fermion  operators $C(\vec{x})$, $C^{\dagger}(\vec{x})$  are  replaced by   $\hat{C}^{\dagger}(\vec{x})$, $\hat{C}(\vec{x})$ (and the Hamiltonian $H$ is replaced by $\hat{H}$). The Hamiltonian  $\hat{H}$ without the condensation energy $\frac{1}{g} |\Delta (\vec{x},t)|^2$ is  expressed in terms  of   the particle-hole   Pauli matrices $\tau_{1}$, $\tau_{2}$ and  $\tau_{3}$. We introduce the two-component spinor  $\hat{\Psi}(\vec{x})=\Big[\hat{C}(\vec{x})  ,\hat{C}^{\dagger}(\vec{x})\Big]^{T}$  and find:
\begin{eqnarray}
&&\hat{H}= \int\,d^2x \hat{\Psi}^{\dagger}(\vec{x})\Big[\tau_{3}\hat{h}_{3}+\tau_{2}\hat{h}_{2}+\tau_{1}\hat{h}_{1}\Big]\hat{\Psi}(\vec{x}),\hspace{0.055in}\frac{1}{2}\vec{\partial}_{x}\theta(\vec{x})\equiv\frac{1}{2}\sum_{l}\vec{\partial}_{x}\varphi_{l}(\vec{x}),
\varphi_{l}(\vec{x})\equiv \arg\Big(\vec{x}-\vec{R}_{l}\Big) , \nonumber\\&&
\hat{h}_{3}=\frac{ v}{2k_{F}}\Big(\Big(-i\vec{\partial}_{x} -\frac{e}{\hbar} \vec{A}(\vec{x})+\frac{1}{2}\vec{\partial}_{x}\theta(\vec{x})\Big)^2- k^2_{F}\Big),\hspace{0.055in} \hat{h}_{2}= i\frac{|\Delta(\vec{x})|}{2k_{0}} \partial_{1},\hspace{0.025in}\hat{h}_{1}= -i\frac{|\Delta(\vec{x})|}{2k_{0}} \partial_{2} . \nonumber\\&&
\end{eqnarray}
The spinor $\hat{\Psi}(\vec{x})=\Big[\hat{C}(\vec{x})  ,\hat{C}^{\dagger}(\vec{x})\Big]^{T}$  \cite{Jackiw}  contains  two parts, the non-zero mode  $\hat{\Psi}_{\neq 0}(\vec{x})$ and the zero mode (Majorana  fermions) $\hat{\Psi}_{0}(\vec{x})$,  $\hat{\Psi}(\vec{x})\equiv \hat{\Psi}_{\neq 0}(\vec{x})+\hat{\Psi}_{0}(\vec{x})$.

\vspace{0.2 in}

\textbf{III. TS Abrikosov vortex lattice}

Next we discuss in detail the non-zero and zero modes of the Abrikosov lattice in a TS. 
\vspace{0.1 in}

\textbf{A. Non-zero modes}

In this section we consider the  non-zero modes for an Abrikosov vortex lattice in the presence of a magnetic field.
The experimental work  on the TS Bi$_{2}$Te$_{3}/$NbSe$_{2}$  \cite{J-P.Xu} shows that an {Abrikosov  vortex lattice} is formed.
 A vortex lattice is stabilized for  superconductors when the penetration depth of the magnetic field is larger than the coherence length \cite{Tinkham,Abrikosov}.
Following  {Appendix A} we find
for a single vortex  a string-like solution for the effective magnetic field. For $|\vec{\hat{x}}|>d\approx \lambda_{L}$ the magnetic field vanishes (d is the vortex lattice constant and $\lambda_{L}$ is the magnetic penetration depth). Since we are interested in the long distance behavior we can approximate the magnetic field for  $|\vec{x}|<\lambda_{L}$  by  a constant field, which is the spatial  average around the vortex core with a radius of $d$. 

Following refs. \cite{Abrikosov,Tinkham} we solve the $p$-wave Hamiltonian in a periodic magnetic field $b$. The periodicity being $d\approx \lambda_{L}$.  The periodic spinor solution is given by:

\noindent
$\mathbf{W(\vec{x})= \sum_{n_{2}} e^{i qn_{2}y} f_{n_{2}}(x) W(x,n_{2})}$, where $\mathbf{ f_{n_{2}}(x) = e^{\frac{-b}{2}(x-\frac{q}{b}n_{2})^2}}$ and $\mathbf{W(x,n_{2}) = \Big[U(x,n_{2}),V(x,n_{2})\Big]^{T}}$ 
 is a two component spinor which is given by the eigenfunction of the $p$-wave Hamiltonian. 
\noindent
 We consider a square lattice and the solution is periodic in the $y$ direction  with the periodicity $d=d_{y}=\frac{2\pi}{q}$. The   periodicity in the  $x$ direction  $d=d_{x}=\frac{q}{b}$ is achieved by demanding   the invariance of the Hamiltonian under the transformation, $n_{2}\rightarrow n_{2}+1$. The system has a finite  extention $L$  in the $x$ direction. Therefore  the value of the momentum in the $y$ direction must be restricted  to $|n_{2}|<n_{max.}=\frac{bL}{q}=\frac{2L}{d}$ (to ensure that the states lie in the box $L\times L$).

Due to the spinor structure of the solution it is convenient to solve the problem in the momentum space and at the end to  impose the periodicity of the wave function.
We represent the  periodic solution as: 
 $\mathbf{W(\vec{x})=\int\,\frac{d^2  k}{(2\pi)^2}e^{i\vec{k}\cdot\vec{x}}W(\vec{k})}$.  
 We use the momentum  representation, $x=i\partial_{k_{1}}$  and find that for the $p$-wave Hamiltonian   $h(\vec{k},i\partial_{k_{1}})$ in the magnetic field $b$:
\begin{eqnarray}
&&h(\vec{k},i\partial_{k_{1}})=\tau_{3}\frac{v}{2k_{F}}\Big[ k^2_{1}+\Big(k_{2}+ib\partial_{k_{1}}\Big)^2-k^2_{F}\Big] +\tau_{1}\frac{\Delta}{2k_{0}}k_{2}-\tau_{2}\frac{\Delta}{2k_{0}}k_{1} , \nonumber\\&&
h(\vec{k},i\partial_{k_{1}})W(\vec{k})=E(\vec{k})W(\vec{k}) , \nonumber\\&&
\end{eqnarray}
where $b i\partial_{k_{1}}$ is the $y$ component of the vector potential in the Landau gauge for the   periodic  magnetic field. 
The eigenvectors and eigenvalues are computed next. The   ground state is given in terms of the ground state energy of the   Harmonic oscillator solution,   $ \epsilon_{0}(b,k_{F})\equiv \frac{vk_{F}}{2}\Big(\sqrt{\frac{b^2}{2}}\Big)$:
\begin{eqnarray}
&&W(\vec{k})=e^{i\frac{k_{1}k_{2}}{b}}e^{-\frac{k^2_{1}\sqrt{2}}{b}}W(\vec{k}),\nonumber\\&&
W(\vec{k})=\frac{1}{\sqrt{2}}\Big[e^{-i\chi(\vec{k})}\sqrt{1+ \frac{\epsilon_{0}(b,k_{F})}{E(\vec{k})}},\sqrt{1- \frac{\epsilon_{0}(b,k_{F})}{E(\vec{k})}}\Big]^{T} , \nonumber\\&&
E(\vec{k})=\sqrt{\epsilon^{2}(b,k_{F})+(\frac {\Delta }{2k_{0}})^2(k^2_{1}+k^2_{2})},\hspace{0.1 in}\epsilon_{0}(b,k_{F})-\frac{v}{2k_{F}}k^{2}_{F} \equiv \epsilon_{0}(b,k_{F})-\frac{vk_{F}}{2}k^2_{F}= \frac{vk_{F}}{2}\Big(\frac{1}{2\pi}\frac{q^2}{k^2_{F}}-1\Big) . \nonumber\\&&
\end{eqnarray}
We note that ${\Delta }/{2k_{0}}$ is  dimensionless.
At this stage we impose the periodic boundary conditions. This results in replacing $k_{1}=qn_{1}$, $k_{2}=qn_{2}$ with the condition  $|n_{2}|<n_{max.}=\frac{bL}{q}=\frac{2L}{d}$, $q=\frac{2\pi}{d}$. 
As a result the eigenspinors and eigenvectors are given  in terms of the integers  $n_{1}, n_{2}$:
\begin{eqnarray}
&&W(n_{1},n_{2})=\sqrt{\frac{1}{2n_{max.}+1}}e^{i\frac{q^2n_{1}n_{2}}{b}}e^{-\frac{ q^2 n^2_{1}\sqrt{2}}{b}}\hat{\hat{W}}(n_{1},n_{2}) , \nonumber\\&&
W(n_{1},n_{2})=\frac{1}{\sqrt{2}}\Big[e^{-i\chi(n_{1},n_{2})}\sqrt{1+ \frac{\epsilon(b,k_{F})}{E(n_{1},n_{2})}},\sqrt{1- \frac{\epsilon(b,k_{F})}{E(n_{1},n_{2})}}\Big]^{T}\equiv \Big[ U(n_{1},n_{2}),V(n_{1},n_{2})\Big] , \nonumber\\&&
E(n_{1},n_{2})=\sqrt{\epsilon^{2}(b,k_{F})+(\frac {\Delta q }{2k_{0}})^2(n^2_{1}+n^2_{2})},\hspace{0.1 in}\epsilon(b,k_{F})=\epsilon(b,k_{F})\equiv\frac{vk_{F}}{2}\Big(\frac{q^2}{8\pi k^{2}_{F}}-1\Big) . \nonumber\\&&
\end{eqnarray}
The spinor $\mathbf{W(n_{1},n_{2})}$  determines the non-zero mode fermion fields.  We introduce  the annihilation and creation operators $\eta(n_{1},n_{2})$,  $\eta^{\dagger}(n_{1},n_{2})$  with respect to the exact ground state $|G\rangle$, $\eta(n_{1},n_{2})|G\rangle=0$  which allows us to write:
\begin{eqnarray}
&&\hat{\Psi}_{\neq 0}(\vec{x})=\sum_{n_{1}}\sum_{n_{2}}e^{i qn_{1} x_{1} +i\frac{q}{2}n_{2} x_{2}}\Big[\eta(n_{1},n_{2})\mathbf{W(n_{1},n_{2})}+\eta^{\dagger}(-n_{1},-n_{2})\tau_{1}\otimes I \mathbf{W^*(-n_{1},-n_{2})}\Big] , \nonumber\\&&
\hat{\Psi}^{\dagger}_{\neq 0}(\vec{x})=\sum_{n_{1}}\sum_{n_{2}}e^{-i qn_{1} x_{1} -i\frac{q}{2}n_{2} x_{2}}\Big[\eta^{\dagger}(n_{1},n_{2}) \mathbf{W^*(n_{1},n_{2})}+\eta(-n_{1},-n_{2})\mathbf{W(n_{1},n_{2})}\tau_{1}\otimes I\Big] . \nonumber\\&&
\end{eqnarray}
We  have replaced the discrete sites  by  the coordinate $x$  in order to consider even and odd rows that we  need to introduce in the matrix $I$ and  double the dimension of   the spinor. (This needs to be done in order to have the same dimension for the zero and nonzero modes). Taking in consideration  also the particle-hole symmetry   we use the  representation $\tau_{1}\otimes I$, which is four-dimensional;  $\tau_{1}$ acts on  the particle-hole space  and $I=1,2$ acts on   the  even and odd  row in real space.

\vspace{0.2 in}

\textbf{B. Zero modes }

\vspace{0.2 in}

In this section we consider the zero modes for an Abrikosov  vortex lattice. In the absence of the vortex lattice 
the zero-mode solutions  for the Hamiltonian in Eq. $(2)$ are given by:  $ W_{0}(\vec{x}) \equiv W_{0}(r,\varphi)=\Big[U_{0}(r,\varphi),V_{0}(r,\varphi)\Big]^{T}\equiv\Big[\frac{1}{\sqrt{i}}e^{\frac{i }{2}\varphi},\frac{1}{\sqrt{-i}}e^{\frac{-i }{2}\varphi}\Big]^{T}\frac{F(r)}{\sqrt{r}}$ (see   {Appendix B})
The function $\frac{F(r)}{\sqrt{r}}$ obeys the normalization condition $\int\,d^2 r [\frac{F(r)}{\sqrt{r}}]^{2}<\infty$.
Due to the charge conjugation property of the Hamiltonian, the zero modes are  Majorana modes.
The solution obtained here is similar to the solution given in refs. \cite{ReadI,Taylor}  for the  $p$-wave superconductors. The explicit form of the kinetic operator   determines  the exact    form of the amplitude $\frac{F(r)}{\sqrt{r}}$  for  the zero modes \cite{DaSarma}.
We consider the case where  the gauged  transformed Hamiltonian  $\hat{H}$  has  $2N$ Majorana zero modes.
The Majorana operators  obey $\gamma_{l}= \gamma^{\dagger}_{l}$, 
$\gamma^{2}_{l}=\frac{1}{2}$ for $l=1, ..., 2N$, 
\begin{equation}
\hat{\Psi}_{0}(\vec{x})=\sum_{l=1}^{2N}\gamma_{l}\Big[\frac{1}{\sqrt{i}}e^{\frac{i }{2}\varphi_{l}(\vec{x})},\frac{1}{\sqrt{-i}}e^{\frac{-i }{2}\varphi_{l}(\vec{x})}\Big]^{T}\frac{F(|\vec{x}-\vec{R}_{l}|)}{\sqrt{|\vec{x}-\vec{R}_{l}|}}  . 
\label{mode}
\end{equation}

 In the second stage we want to discuss the effect of the vortex  lattice  on the localized    Majorana fermions. The effect of the vortex lattice is to delocalize the Majorana fermions and form  dispersive Majorana  bands.
This is a result  due to ref. \cite{Stern} which showed that Majorana fermions enclose a flux which depends on the number of vortices on a closed polygon. The flux on a polygon  of $n$ vortices is $\frac{\pi}{2}(n-2) $ (n is the number of vortices in the polygon) \cite{Stern}.
For a  square vortex lattice with four  vortices ($n=4$) per plaquette the flux will be  $\frac{\pi}{2}(4-2)=\pi $.  
 For the Majorana case we restrict ourselves  to plaquettes  with four vortices (per plaquette) \cite{Stern}.  We consider the effect of the overlap between Majorana fermions given by matrix element  $t_{0}$   \cite{Stern,Biswas}. 
\begin{equation}
H_0 = i t_{0}\sum_{i,j}\hat{\Psi}_{0}^{\dagger}(x_{i})S(x_{i},x_{j})\hat{\Psi}_{0}(x_{j}) ,  
\label{equation}
\end{equation}
where $S(x_{i},x_{j})$ introduces  the phase on the bond $i,j$ and determines the flux of $\pi$ per plaquette and  $t_{0}$ is the overlap between the Majorana fermions which is  determined from Eq. $(7)$. 
(The minimum  energy  for  the Abrikosov vortices  is obtained  for  a triangular lattice \cite{Abrikosov}. For  a square lattice   the energy is less favorable, but it is simpler to analyze.)

We   choose   a gauge for which the hopping constant along columns has  positive sign  and alternating signs between adjacent rows.
This means that $S(x_{i},x_{j})=|S(x_{i},x_{j})|e^{i\theta_{i,j}}$,   
 the phase $ \theta_{i,j}$ is zero   along columns, has  positive sign ($\theta_{i,j}=0$)  and alternating signs between adjacent rows  ($\theta_{i,j}=\pi$) \cite{Stern}.  As a result  we obtain two flat  bands  and a third band that is dispersive and gapless \cite{Biswas,Stern,Franz} comprising  Majorana fermions with the eigenvalues: 
\begin{equation}
\lambda^{0}_{\alpha=1}(\vec{k})\equiv \lambda(\vec{k})= t_{0}\sqrt{\sin^2(k_{1})+\sin^2(k_{2})}, \hspace{0.1in }|\vec{k}|\leq{\frac{\pi}{2}},\hspace{0.1 in}\lambda^{0}_{\alpha=2}(\vec{k})=0 . 
\label{zerovector}
\end{equation}
 For the zero modes we find that the  representation of the zero modes is given  in terms of the zero mode Majorana operators  in the continuum representation  $\Gamma(\vec{x})=\Gamma^{\dagger}(\vec{x})$   ($x_{i}$ is replaced  with the continuum coordinate  $x$,  even and odd rows are introduced with the help of the   matrix $I$) and spinor  eigenfunctions are given in momentum space $ L(\vec{k})$:
\begin{eqnarray}
&&\hat{\Psi}_{0}(\vec{x})=\sum_{k}e^{i\vec{k}\cdot \vec{x}}\Big[\Gamma(\vec{k})L(\vec{k})+\Gamma^{\dagger}(-\vec{k})\tau_{1}\otimes IL^{*}(-\vec{k})\Big] , \nonumber\\&&
\hat{\Psi}^{\dagger}_{ 0}(\vec{x})=\sum_{k}e^{-i\vec{k}\cdot \vec{x}}\Big[\Gamma^{\dagger}(\vec{k})L^*(\vec{k})+\Gamma(-\vec{k})L(-\vec{k})\tau_{1}\otimes I\Big] . \nonumber\\&&
\end{eqnarray}
The four  components of the spinor  $L(\vec{k})$ in Eq. $(10)$ are given by:
\begin{eqnarray}
&&L(\vec{k})=  \frac{1}{\sqrt{2}\sqrt{\sin^2[k_{1}]+4(\sin[k_{1}]+\lambda(\vec{k}))^2}}  \nonumber\\&&\Big[\sin[k_{2}], \sin[k_{2}], \frac{-i (\sin[k_{1}]+\lambda(\vec{k}))}{\sin[ k_{2}]}(1-e^{2ik_{2}}), \frac{-i (\sin[k_{1}]+\lambda(\vec{k}))}{\sin[ k_{2}]}(1-e^{2ik_{2}})\Big]^{T} . \nonumber\\&&
\end{eqnarray}
 We replace the discrete sites $x_{i}$  by  the coordinate $x$, therefore we introduce the matrix $I$ to double the dimension of   the spinor.  Taking in consideration  also the particle-hole symmetry   we   use the  representation $\tau_{1}\otimes I$ which is four-dimensional;  $\tau_{1}$ acts on  the particle-hole space  and $I=1,2$ acts on   the  even and odd  row.
We note that  a   triangular lattice  for the Majorana modes   has been considered in refs. \cite{Biswas,Stern}.


\vspace{0.2 in}

\textbf{IV. Viscosity tensor for  the  TS } 

\vspace{0.2 in}

\noindent

In the first part we  introduced the theory for the dissipative viscosity and in the second part we  used this theory to investigate  the  Abrikosov lattice. 
The physics of  solids  \cite{LandauI} and fluids \cite{LandauII} provides  us the relation  between  the stress tensor $ \Sigma_{i,j}$, strain field  $\epsilon_{kl}$  and the velocity strain  field $v_{kl}=\partial_{t}(\epsilon_{kl})$  \cite{Kosevich}.
We now use  this description    for   quantum fluids in a solid.
The combination of the stress tensor resulting from a strain field and  the dissipative part of the stress determines  the equation:
\begin{equation}
\Sigma_{i,j}=\lambda_{ij,kl}\epsilon_{kl}+\zeta_{ij,kl}v_{kl} . 
\label{relation}
\end{equation}
The strain field $\epsilon_{k,l}$  is given in terms of the lattice deformation  $\vec{u}(\vec{x},t)$, $\epsilon_{k,l}=\frac{1}{2}(\partial_{k}u^{l}+\partial_{l}u^{k})$, $\epsilon_{0,i}= \partial_{t}u^{i}$. 
The viscosity tensor is  given by $\zeta_{ij,kl}\equiv\frac{\Sigma_{i,j}}{v_{k,l}}$, and 
 $\zeta_{ij,kl}$   is separated into two parts, $\zeta_{ij,kl}=\zeta^{S}_{ij,kl}+\zeta^{A}_{ij,kl}$,     $\zeta^{S}_{ij,kl}=\zeta^{S}_{kl,ij}$ is  the  symmetric  part  and     $\zeta^{A}_{ij,kl}=-\zeta^{A}_{kl,ij}$ is the antisymmetric part \cite{Avron}.
From the Onsager relations \cite{Onsager} we know that when the  time reversal   symmetry is violated, like in the quantum Hall system  and the $p$-wave superconductor case \cite{Avron,Yakovenko,Read,Son},  we have $\zeta^{A}_{ij,kl}\neq 0$.
 For  $\zeta_{ii,kk}$ with  $i\neq k$ we have  a situation  where the  strain  field generates  stress  in the  perpendicular direction. 
 The stress tensor for quantum fluids  is   obtained from  the invariance of the Lagrangian under an arbitrary  local coordinate transformation \cite{Nakahara,Katanaev,DavidR,DavidD}.  The  explicit dependence of the strain field on the lattice deformation  determines the coordinate transformation \cite{Katanaev} from which we obtain the stress tensor. In the presence of an elastic deformation $\vec{u}$ the coordinates  transform in the following way: $ \vec{x}\rightarrow \vec{\xi}(\vec{x})= \vec{x}+\vec{u}(\vec{\xi})$ \cite{Katanaev} (see {Appendix C}).

This coordinate transformation  allows the identification  of the stress tensor.
Using the invariance  of the {\it spinless} fields   under the coordinate deformation   gives,  
\begin{equation}
\mathbf{\hat{C}}(\vec{\xi}(\vec{x}))=\hat{C}(\vec{x}) ,\hspace{0.15 in} \mathbf{\hat{C}^{\dagger}}(\vec{\xi}(\vec{x}))=\hat{C}^{\dagger}(\vec{x}).
\label{fields}
\end{equation}
For the   $p$-wave Hamiltonian  given  in Eq. $(2)$  we find 
  the momentum density $\pi_{i}(\vec{x},t)$ and stress tensor  $\Sigma_{ij}(\vec{x},t)$ induced by the strain fields $\epsilon_{0,i}(\vec{x},t)$, $\epsilon_{i,j}(\vec{x},t)$. To linear order in the deformation strain field  we obtain the response stress field. Using the invariance, Eq. $(13)$, with respect to the coordinate transformation  determines  the strain-stress Hamiltonian   $H^{ext.}(t)$:
\begin{eqnarray}
&& H^{ext.}(t)=\int\,d^{2}x\Big[\epsilon_{0,1}(\vec{x},t)\pi_{1}(\vec{x},t)
+\epsilon_{0,2}(\vec{x},t)\pi_{2}(\vec{x},t) +\epsilon_{1,1}(\vec{x},t) \Sigma_{11}(\vec{x},t) +\epsilon_{22}(\vec{x},t) \Sigma_{22}(\vec{x},t) \nonumber\\&&
+\epsilon_{12}(\vec{x},t) \Sigma_{12}(\vec{x},t)  +\Omega_{12}(\vec{x},t) R_{12}(\vec{x},t) \Big].\nonumber\\&&
\end{eqnarray}
In the absence of disclinations $\Omega_{12}(\vec{x},t)=0$  in Eq. $(14)$.
\noindent

Using the explicit dependence of  the  spinor   in terms of the zero and nonzero modes given in Eqs. (5), (10) allows us to represent  the stress fields:
\begin{eqnarray}
&&\pi_{1}(\vec{x},t)=\hat{\Psi}^{\dagger}(\vec{x},t)\Big(I i\partial_{1}\Big)\hat{\Psi}(\vec{x},t);\hspace{0.1 in}  \pi_{2}(\vec{x},t)=\hat{\Psi}^{\dagger}(\vec{x},t)\Big(I i\partial_{2}\Big)\hat{\Psi}(\vec{x},t),\nonumber\\&&
\Sigma_{11}(\vec{x},t)\approx -\frac{\Delta}{2k_{0}}\hat{\Psi}^{\dagger}(\vec{x},t)\Big(\tau_{2}(-\partial_{1})\hat{\Psi}(\vec{x},t)\Big); ~~
\Sigma_{22}(\vec{x},t)\approx-\frac{\Delta}{2k_{0}}\hat{\Psi}^{\dagger}(\vec{x},t)\Big(\tau_{1}(-i\partial_{2})\Big)\hat{\Psi}(\vec{x},t),\nonumber\\&&
\Sigma_{12}(\vec{x},t)\approx \frac{\Delta}{2k_{0}}\hat{\Psi}^{\dagger}(\vec{x},t)\Big(\tau_{2}(-i\partial_{2})+\tau_{1}(-i\partial_{1})\Big)\hat{\Psi}(\vec{x},t); \nonumber\\&&
R_{12}(\vec{x},t)\approx   \frac{\Delta}{2k_{0}}\hat{\Psi}^{\dagger}(\vec{x},t)\Big(\tau_{2}(-i\partial_{2})-\tau_{1}(-i\partial_{1})\Big)\hat{\Psi}(\vec{x},t);  ~~
\hat{\Psi}(\vec{x})\equiv \hat{\Psi}_{\neq 0}(\vec{x})+\hat{\Psi}_{0}(\vec{x}) . \nonumber\\&&
\end{eqnarray}

\noindent
  We  compute the viscosity  stress tensor   and the {\it dissipative}   viscosity tensor $\zeta_{ij,kl}(\vec{q},\Omega)=\frac{\Sigma_{ij}(\vec{q},\Omega)}{v_{kl}(-\vec{q},-\Omega)}$ in quantum fluids.
\noindent 
 We vary the $p$-wave  Hamiltonian  by a  linear   strain field $\delta u^{i}(\vec{x},t)$,  $ u^{i}(\vec{x},t)\rightarrow u^{i}(\vec{x},t)+\delta u^{i}(\vec{x},t)$ and  find from Eq. $(6)$ that the variation with respect to $\delta u^{i}(\vec{x},t)$ satisfies  the  continuity equation:
\begin{equation}
\partial_{t}\pi_{j}(\vec{x},t)+\sum_{i}\partial_{i}\Sigma_{i j}(\vec{x},t)=0.
\label{conservation}
\end{equation}
\noindent
Using   the linear response theory \cite{Fetter}   with respect to the   strain-stress Hamiltonian   $H^{ext.}(t)$  given in Eq. $(14)$ we obtain:
\begin{equation}
\langle G|\Sigma_{ij}(x,t)|G\rangle_{ext.}=  \langle G|\Sigma_{ij}(x,t)|G\rangle +(\frac{-i}{\hbar})\int_{0}^{t}\,dt'\langle G|[\Sigma^{H}_{ij}(x,t), H_{H}^{ext.}(t')]|G\rangle,
\label{eq}
\end{equation}
where $\Sigma^{H}_{ij}(x,t)$ is the stress in the Heisenberg representation with  the  ground state $|G\rangle$ given in Eqs. (6), (11).
 Following ref. \cite{Fetter} we obtain the relation $\Sigma_{ij}(\vec{q},\Omega)=\zeta_{ij,kl}(\vec{q},\Omega)v_{kl}(-\vec{q},-\Omega)$: 
\begin{eqnarray}
&&R_{ij,kl}(\vec{q},t)=(\frac{-i}{\hbar})\langle G|\mathbf{T}\Big(\Sigma_{ij}(\vec{q},t)\Sigma_{kl}(-\vec{q},0)\Big)|G\rangle,  ~~ R_{ij,kl}(\vec{q},\Omega)=\int_{-\infty}^{\infty}R_{ij,kl}(\vec{q},t)e^{-i\Omega t}\,dt,\nonumber\\&&
\zeta_{ij,kl}(\vec{q},\Omega)=-\frac{R_{ij,kl}(\vec{q},\Omega)}{i\Omega}.\nonumber\\&&
\end{eqnarray}
The {\it dissipative}  $\zeta_{ij,kl}(\vec{q},\Omega)$ part of the  viscosity tensor is obtained after the analytic continuation $i\Omega\rightarrow \Omega+i0^{+}$. Equation $(18)$ is computed  using Wick's  theorem \cite{Fetter} for  the stress $\Sigma_{ij}(\vec{q},t)$  which is expressed in terms of  the spinor   $\hat{\Psi}_(\vec{x})=\hat{\Psi}_{\neq 0}(\vec{x})+\hat{\Psi}_{ 0}(\vec{x})$ in Eq. $(15)$.

\vspace{0.2 in}

\textbf{V. Application of the  viscosity tensor to  ultrasound attenuation} 

\vspace{0.2 in}
In this section we compare our  calculation  to the existing   ultrasound attenuation method  given in the literature \cite{Tinkham,Schrieffer,Philip}.  For  the  topological superconductors we need to work with   the   spinor $\hat{\Psi}(\vec{x})=\hat{\Psi}_{\neq 0}(\vec{x})+\hat{\Psi}_{ 0}(\vec{x})$.  Due to the dispersive nature of the  zero modes,  we will have new contributions to the absorption from the mixed  pairs $\hat{\Psi}_{\neq 0}(\vec{x})$ and $\hat{\Psi}_{ 0}(\vec{x})$.

In a superconductor the electron-phonon interaction  couples to longitudinal as well as transverse phonons. The coupling of the electrons to  the transverse phonons in the superconducting phase  is less understood.  We show that by applying a transverse strain we can obtain non-diagonal response for stress; in this way we study the transverse effect of phonons.

The ultrasound attenuation method measures the superconducting gap.
The single particle  contribution  to the absorption  in a superconductor    is given by $\frac{\alpha_{sc}}{\alpha_{normal}}=\frac{2}{e^{\frac{\Delta(T)}{T}} +1}$  ($\alpha_{sc}$ is the absorption for the superconductor and $\alpha_{normal}$  is the absorption  in the normal phase). In our case we have contributions from the electrons and the Majorana modes. Due to the magnetic vortex lattice the absorption is given by discrete summations instead of the integration of quasi-particle density of states.
The absorption of  transverse phonons is  obtained from the response of  non-diagonal strain tensor $\epsilon_{ij}(\vec{x},t)$.
 Theoretically  we express the strain tensor $\epsilon_{ij}(\vec{x},t)$ in terms of the normal modes  of the harmonic crystal \cite{Kosevich}  [$\epsilon_{ij}(\vec{x},t)$  depends only on  the   crystal   phonons]  without requiring knowledge of the explicit  electron-phonon interaction.
 The strain tensor   $\epsilon_{ij}(\vec{x},t)$ is represented  in terms   of  the normal phonon operators  $b_{s}(\vec{Q})$ and  $b^{\dagger}_{s}(\vec{Q})$ ($s=1,2$ are the two phonon polarizations for a phonon in the $i$ direction  given by the vector $\vec{e}$ in the orthogonal direction to the  vector $\vec{Q}$):  
\begin{equation}
\epsilon_{ij}(-\vec{Q},t)=\sum_{s}\frac{1}{2\pi}\sqrt{\frac{\hbar}{2\rho}}\frac{1}{\sqrt{\omega_{s}(\vec{Q})}}\Big[ie^{(i)}(\vec{Q},s)Q_{j}(b^{\dagger}_{s}(\vec{Q})+b_{s}(-\vec{Q}))\Big] . 
\label{normal}
\end{equation}
We substitute Eq.  (19) into Eq. (14) with  the representation of  the stress tensor given in   Eq. (15). Using first order perturbation theory we  compute the ultrasound attenuation  in agreement with  refs. \cite{Tinkham,Schrieffer,Philip}.

 Following ref. \cite{LandauI},  the transverse sound absorption $\alpha_{t}(\Omega)$ and longitudinal absorption  $\alpha_{t}(\Omega)$ can be represented  in terms of the viscosity tensor $\zeta_{12,12}(\vec{q},\Omega)$ and  $\zeta_{11,11}(\vec{q},\Omega)$ defined in Eq. (18).   In two dimensions we have  for the transverse absorption,
$\alpha_{t}(\Omega)=\zeta^{\perp}(\Omega)\frac{\Omega^2}{2 \rho v^{3}_{s,\perp}}$ where $\zeta^{\perp}(\Omega)=\frac{1}{2}\Big(\zeta_{12,12}(\Omega)+\zeta_{21,21}(\Omega)\Big)$.
The longitudinal absorption is given by $\alpha_{l}(\Omega)=\zeta^{\parallel}(\Omega)\frac{\Omega^2}{2 \rho v^{3}_{s,\parallel}}$  where $\zeta^{\parallel}(\Omega)=\frac{1}{2}\Big(\zeta_{11,11}(\Omega)+\zeta_{22,22}(\Omega)\Big)$.
\noindent
 The information about the crystal enters through the sound velocity $v_{s,\perp}$ (transverse), $v_{s,\parallel}$ (longitudinal), crystal density $\rho$ and quantum fluid viscosity   $\zeta_{ij,kl}(\vec{q},\Omega)$.
For example the viscosity terms  $\zeta_{11,11}(\vec{q},\Omega)$ are  computed according to Eq. (18) using the mode expansion of  the spinor  $\hat{\Psi}(\vec{x})=\hat{\Psi}_{\neq 0}(\vec{x})+\hat{\Psi}_{ 0}(\vec{x})$ which allows us to compute the longitudinal  absorption:
$R^{\neq 0,\neq 0}_{||,||}(\vec{Q},t)$  represents  the non-zero mode part  [the index ${\neq 0,\neq 0}$ means that the two fields which contribute  to  absorption  are  only non-zero modes, the symbol ${||,||}$ means  that we have contributions only from $\zeta_{11,11}(\Omega)$ and $\zeta_{22,22}(\Omega)$].
Similarly, $R^{\neq0,0}(\vec{Q},t)$ represents the mixed contribution, a zero mode and a non-zero mode (one field contains the zero mode and the second contains the non-zero mode) whereas  $R^{ 0,0}(\vec{Q},t)$  represents the contribution when both fields are  zero modes.

Using the imaginary time order operator $ T_{t}$  in the imaginary  time representation \cite{Fetter} we find:
for $ i=2$ the tensor $R_{22,22}(\vec{Q},t)$ is given as $R_{22,22}(\vec{Q},t)=-T_{t}\langle G|\Sigma_{22}(\vec{Q},t)\Sigma_{22}(-\vec{Q},t)|G\rangle$, and 
$R_{22,22}(\vec{Q},t)=R^{\neq 0,\neq 0}_{22,22}(\vec{Q},t)+R^{\neq 0,0}(\vec{Q},t)+  R^{0,\neq0}(\vec{Q},t)+R^{0,0}(\vec{Q},t)$.

\noindent

The  {\it particle-hole} contribution is given by  $R^{\neq 0,\neq 0}_{22,22}(\vec{Q},t;qp)$ (the particle-particle contributions are neglected and the symbol $qp$ means particle-hole).
The mixed terms particle-Majorana and hole-Majorana are given by  
 $R^{\neq 0,0}(\vec{Q},t;qp)+  R^{0,\neq0}(\vec{Q},t;qp)$.
Using Wick's theorem with the  spinor representation  given in Eqs. (5), (10) we  compute   $R_{22,22}(\vec{Q},t)$ and $\zeta_{22,22}(\vec{Q},\Omega)$.

\textbf{a. Non-zero modes  absorption}

We consider first the absorption for the particle-hole in the absence of  Majorana modes. We find for the dissipative  viscosity $\zeta_{22,22}(\vec{Q},\Omega;qp)$:

\begin{eqnarray} 
&&\zeta_{22,22}(\vec{Q},\Omega;qp)=(\frac{\Delta}{k_{0}})^2\Big[\frac{1}{2}k^{2}_{F}+Q_{2}^2\Big]\Big[\sum_{n_{1}}\sum_{n_{2}}
\Big[\frac{1}{e^{\frac{E(n_{1},n_{2})}{T}}+1}-\frac{1}{e^{\frac{E(n_{1},n_{2}-\frac{Q_{2}d}{2\pi})}{T}}+1}\Big]\nonumber\\&&\Big(\frac{-i}{\Omega}\Big)\Big[\frac{1}{i\hat{\Gamma}+\Omega+E(n_{1},n_{2})-E(n_{1},n_{2}-\frac{Q_{2}d}{2\pi })}-\frac{1}{i\hat{\Gamma}+\Omega-E(n_{1},n_{2})+E(n_{1},n_{2}-\frac{Q_{2}d}{2\pi })}\Big] , 
\nonumber\\&&
\end{eqnarray}
where $ E(n_{1},n_{2})$ is the quasi-particle dispersion given in Eq.  (11). Here $\hat{\Gamma}$ denotes  the   scattering life-time for  the quasi-particles,    $d$ is the lattice separation between the vortices, $\Omega$ is the frequency of the transducer  strain field  and $T$ is the temperature.

For an  $s$-wave superconductor  the  absorption  agrees with the results  given in the literature, $\frac{\alpha_{sc.}}{\alpha_{normal}}=\frac{2}{e^{\frac{\Delta(T)}{T}} +1}$.
For the present  case with the dispersion  $ E(n_{1},n_{2})$, the absorption is controlled by the magnetic field with the ground state energy  $\epsilon(b,k_{F})$ [see Eq. $(11)$].

\textbf{b. Absorption due to Majorana modes}

The Majorana modes give rise to the  particle-hole (Majorana) absorption  $\zeta _{22,22}(\vec{Q},\Omega;0,\neq 0;qp)$ and the  particle-particle (Majorana) absorption  $\zeta_{22,22}(\vec{Q},\Omega;0,\neq 0;pp)$. This notation means that the absorption is controlled by {a zero and a non-zero mode, $``0,\neq 0"$} and the non-zero mode is either a {particle-hole $``qp"$} or a {particle-particle  $``pp"$}  channel. Both absorptions are controlled  by the tunneling amplitude $t_{0}$ of the dispersive Majorana mode. 
From the  particle-hole (Majorana) absorption   $\zeta_{22,22}(\vec{Q},\Omega;0,\neq 0;qp)$ we have the combination where the nonzero mode is a particle and the Majorana operator in the fermionic representation  $\Gamma$ [see Eq. $(5)$] is a hole or vice versa. We introduce the scattering life time   ($\hat{\Gamma}$) and find for 
 $\zeta_{22,22}(\vec{Q},\Omega;0,\neq 0;qp)$  the representation:
\begin{eqnarray}
&&\zeta_{22,22}(\vec{Q},\Omega;0,\neq 0;qp)=(\frac{\Delta}{k_{0}})^2\Big[\frac{1}{2}k^{2}_{F}+Q_{2}^2\Big]\Big[\sum_{n_{1}}\sum_{n_{2}}
\Big[\frac{1}{e^{\frac{\lambda(n_{1},n_{2})}{T}}+1}-\frac{1}{e^{\frac{E(n_{1},n_{2}-\frac{Q_{2}d}{2\pi })}{T}}+1}\Big]\nonumber\\&&\Big(\frac{-i}{\Omega}\Big)\Big[\frac{1}{i\hat{\Gamma}+\Omega+\lambda(n_{1},n_{2})-E(n_{1},n_{2}-\frac{Q_{2}d}{2\pi})}-\frac{1}{i\hat{\Gamma}+\Omega-\lambda(n_{1},n_{2})+E(n_{1},n_{2}-\frac{Q_{2}d}{2\pi })}\Big] , 
\nonumber\\&&
\end{eqnarray}
where $ \lambda(n_{1},n_{2})$ is the dispersion of the Majorana fermions given in Eq. (4).

 The particle-particle (Majorana) absorption (this is the case that a non-zero mode particle  and a zero mode Majorana are created)    $\zeta_{22,22}(\vec{Q},\Omega;0,\neq 0;pp)$ is given by:


\begin{eqnarray}
&&\zeta_{22,22}(\vec{Q},\Omega;0,\neq 0;pp)=(\frac{\Delta}{k_{0}})^2\Big[\frac{1}{2}k^{2}_{F}+Q_{2}^2\Big]\Big[\sum_{n_{1}}\sum_{n_{2}}
\Big[1-\frac{1}{e^{\frac{\lambda(n_{1},n_{2})}{T}}+1}-\frac{1}{e^{\frac{E(n_{1},n_{2}-\frac{Q_{2}d}{2\pi })}{T}}+1}\Big]\nonumber\\&&\Big(\frac{\Gamma}{\Omega}\Big)\Big[\frac{1}{(\Omega-\lambda(n_{1},n_{2})-E(n_{1},n_{2}-\frac{Q_{2}d}{2\pi}))^2+\hat{\Gamma}^2}
\Big] , 
\nonumber\\&&
\end{eqnarray}
where $\lambda(n_{1},n_{2})$ is the dispersion of the Majorana fermions given in Eq. $(4)$.
From the theory of  electromagnetic paramagnetic  response in superconductors [see ref. \cite{Schrieffer}, Eqs. (8.47) - (8.50)] we can see the similarity with our results, Eqs. (21), (22). In our case  the electric field is replaced by the strain velocity $\partial_{t}\epsilon_{k,l}$.

\vspace{0.2 in}

\textbf{VI. Transverse impedance  for topological superconductors }

\vspace{0.2 in}

\noindent
The transverse impedance $\zeta_{22,01}(\vec{Q},\Omega;)$ is given by: 
$\zeta_{22,01}(\vec{Q},t)=\frac{\Sigma_{22}(\vec{Q},\Omega;)}{\partial_{t}u^{1}(t)}$.  
It  describes the response of the stress field in the $i=2$ direction  to an applied  strain field in the $i=1$ direction similar to the Hall effect.
In the frequency space we have: 
\begin{eqnarray}
&&\Sigma_{22}(\vec{Q},\Omega)=\Big[\frac{\lambda_{22,01}}{i\Omega}+\zeta_{22,01}(\vec{Q},\Omega;)\Big ]v_{1}(\Omega) ,\hspace{0.1 in} v_{1}(\Omega)=-i\Omega u^{1}(\Omega) , \nonumber\\&&
\zeta_{22,01}(\vec{Q},\Omega;)\equiv \zeta^{R}_{22,01}(\vec{Q},\Omega;)+i\zeta^{I}_{22,01}(\vec{Q},\Omega;) ,\nonumber\\&&
\end{eqnarray}
where  $\zeta^{R}_{22,01}(\vec{Q},\Omega;)$ is the real  dissipative part and $ i\zeta^{I}_{22,01}(\vec{Q},\Omega;)$ is the imaginary part. 
We now compute the  transverse  impedance for the TS Abrikosov  vortex lattice. 

 Similar studies have been performed  for  the    superfluid Helium $^{3}$He phase.   The authors in ref.  \cite {Nagai} have measured  the  superfluid acoustic impedance  of $^{3}$He-B   coated with a wall of several layers of $^{4}$He.  The measurement has been  performed  using the resonance frequency of an ac-cut transducer which oscillates in a shear or longitudinal  mode.  The coating was used to enhance the specularity of quasi-particle scattering  by the wall.  In our case we do not have a rough wall  and do not approximate the scattering by  quasi-classical theory with a random $S$-matrix. Instead, we use  an oscillating wall and compute the dissipative viscosity using the linear response theory given by  Eqs. (14-18). We make use of the full spectrum of the zero and non-zero modes given
in Eqs. (5) - (11).

 \textbf{a. Impedance for non-zero modes}

 The dissipative  quasi-particle contribution  $\zeta^{(R)}_{22,01}(\vec{Q},\Omega;qp)$ in the absence of Majorana modes  is:
\begin{eqnarray} 
&&\zeta^{(R)}_{22,01}(\vec{Q},\Omega;qp)\equiv\frac{ \Delta}{2k_{0}}Q_{1}Q_{2}\Big[\sum_{n_{1}}\sum_{n_{2}}
\Big[\frac{1}{e^{\frac{E(n_{1},n_{2})}{T}}+1}-\frac{1}{e^{\frac{E(n_{1},n_{2}-\frac{Q_{2}d}{2\pi})}{T}}+1}\Big]\nonumber\\&&\Big(\frac{-i}{\Omega}\Big)\Big[\frac{1}{i\hat{\Gamma}+\Omega+E(n_{1},n_{2})-E(n_{1},n_{2}-\frac{Q_{2}d}{2\pi })}-\frac{1}{i\hat{\Gamma}+\Omega-E(n_{1},n_{2})+E(n_{1},n_{2}-\frac{Q_{2}d}{2\pi })}\Big] . 
\nonumber\\&&
\end{eqnarray}
Here  $\Delta$ is  the TS  pairing field, $q=\frac{2\pi}{d} $ is the momentum of the vortex lattice with the vortex  separation distance  $d$, $\Omega$ is the frequency of the applied strain field and  $\hat{\Gamma}$ is the scattering life-time.
We find   $\zeta^{(R)}_{22,01}(\vec{Q},\Omega;qp)\approx  
\frac{\Delta}{2k_{0}} Q_{1}Q_{2}\cdot  10^{5}  I_{qp}(t)\Big[\frac{Newton\cdot sec.}
{m^3}\Big]$.

The gap parameter  $\frac{\Delta}{2k_{0}}$  controls   the dissipative stress.  Using Eq. $(24)$  with the gap parameter   $\frac{\Delta}{2k_{0}}q n_{max}\approx$ 0.1 mev,  momentum    $q=10^{8}$ m$^{-1}$, and  vortex lattice $d=10^{-7}$ m we  have $\frac{\Delta}{2k_{0}}  \approx 10^{-31}$ Joule $\times$ m  and  find: $ \zeta^{(R)}_{22,01}(\vec{Q},\Omega;qp)\approx\frac{Q_{1}}{q}\frac{Q_{2}}{q}q^2 \times 10^{-31}\times  10^5 I_{qp} \Big[\frac{Newton\sec.}
{m^3}\Big]=\frac{Q_{1}}{q}\frac{Q_{2}}{q}10^{-10} I_{qp} \Big[\frac{Newton\cdot sec.}
{m^3}\Big]$, with  $I_{qp}(t)$  varying from  $1$ to $6$.

\begin{figure}
\begin{center}
\includegraphics[width=3.5 in]{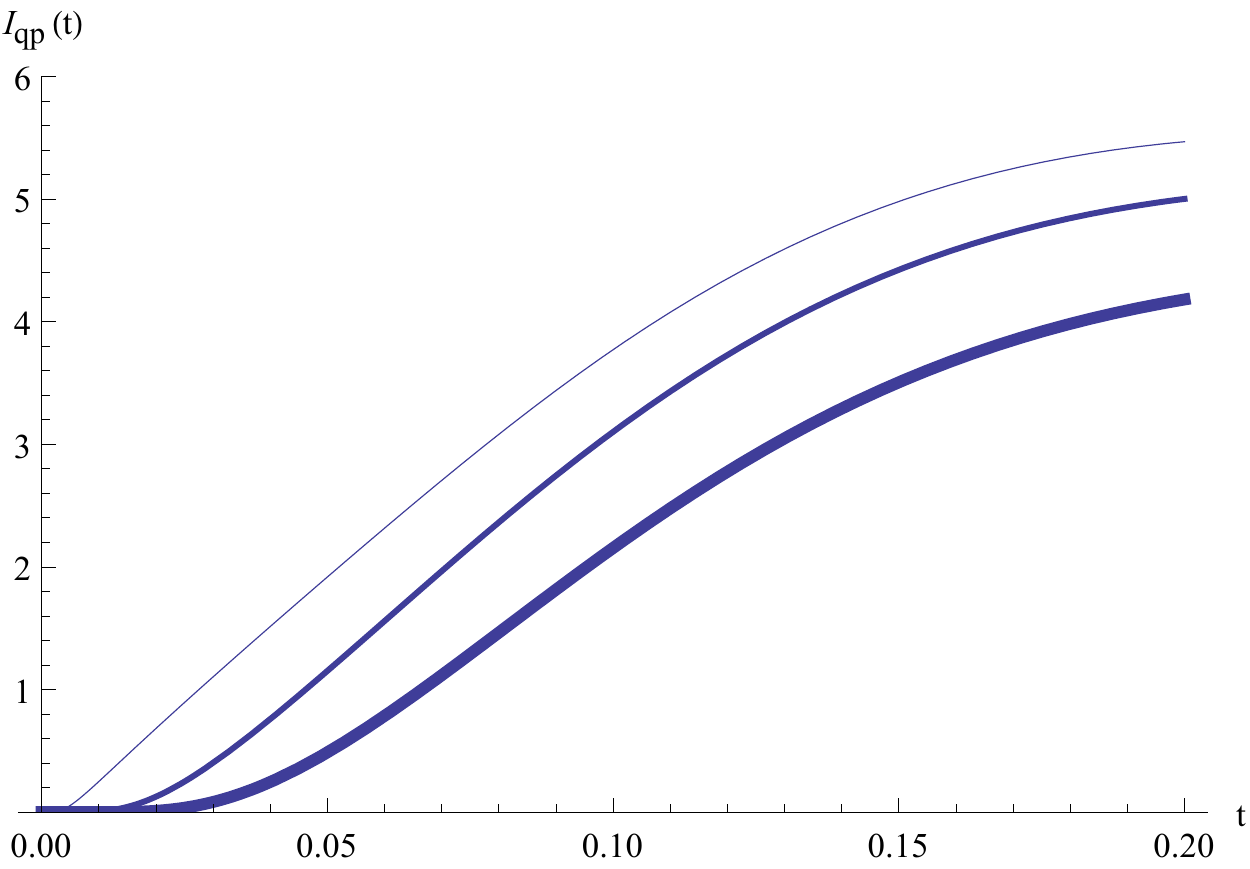}
\end{center}
\caption{The dissipative part  for  the particle-hole  contribution,
$\frac{\zeta^{(R)}_{22,01}(\vec{Q},\Omega;qp)}{10^{-10}}\approx  (\frac{Q_{1}}{q})(\frac{Q_{2}}{q} )I_{pq}(t)\Big[\frac{Newton\cdot sec.}
{m^3}\Big]$, with $I_{qp}(t)$ shown in Fig. 2.}
\end{figure}

Figure $1$ shows that the sound  dissipative impedance  is controlled by the absorption edge condition  $\epsilon(b,k_{F})>t$ [here $t$ is the temperature  and $\epsilon(b,k_{F})$ is the ground state energy  determined by the magnetic field $b$]. Using the explicit formula $ \epsilon(b,k_{F})=\epsilon_{0}(b,k_{F})-\frac{v}{2k_{F}}k^{2}_{F} \equiv  \frac{vk_{F}}{2}\Big(\frac{1}{2\pi}\frac{q^2}{k^2_{F}}-1\Big)$, $\epsilon_{0}(b,k_{F})=\frac{vk_{F}}{2}\sqrt{\frac{b^2}{2}}$  given in Eq. $(11)$ we can determine from the impedance  $\zeta^{(R)}_{22,01}(\vec{Q},\Omega;qp)$ the magnetic field $b$ and  the vortex lattice constant $d$.

The absorption is given in units of $\frac{Newton \cdot sec.}{m^3 }$. We plot $\frac{\zeta^{(R)}_{22,01}(\vec{Q},\Omega;qp)}{10^{-10}}$ for the case $\Delta=0.1$ meV and $\Omega=2\pi  10^{6}  \frac{radians}{sec.}$  as a function of temperature $t$ ($t=0.1$ meV corresponds to $1$ Kelvin)  and the  vortex separation is $d=10^{-7} $ m. The function $I_{qp}(t)$ is shown for three different cases: The {\it thin line}  gives the absorption for the  ground state  energy $\epsilon(b,k_{F})\equiv\frac{vk_{F}}{2}\Big(\frac{q^2}{8\pi k^{2}_{F}}-1\Big)=0.001$ mev, the {\it thickest line} represents the absorption for the  ground state  energy $\epsilon(b,k_{F})=0.1$ meV. The {\it line in between} describes the situation  for   $\epsilon(b,k_{F})=0.0075$ meV.
We observe  that  the absorption  edge  temperature scales with  the energy $\epsilon(b,k_{F})$ as a function of the magnetic field and  Fermi energy.

 \textbf{b. Impedance for Majorana modes }

 The Majorana contribution given by the particle-hole  [see Eq. $(21)$] depends on the tunneling amplitude  $t_{0}$. We consider  the ground state energies  $\epsilon(b,k_{F})=0.005$ meV (thin line),  $\epsilon(b,k_{F})=0.1$ meV (thick line) and   $\epsilon(b,k_{F})=0.05$ meV (intermediate ground state energy).
\begin{figure}
\begin{center}
\includegraphics[width=3.5 in ]{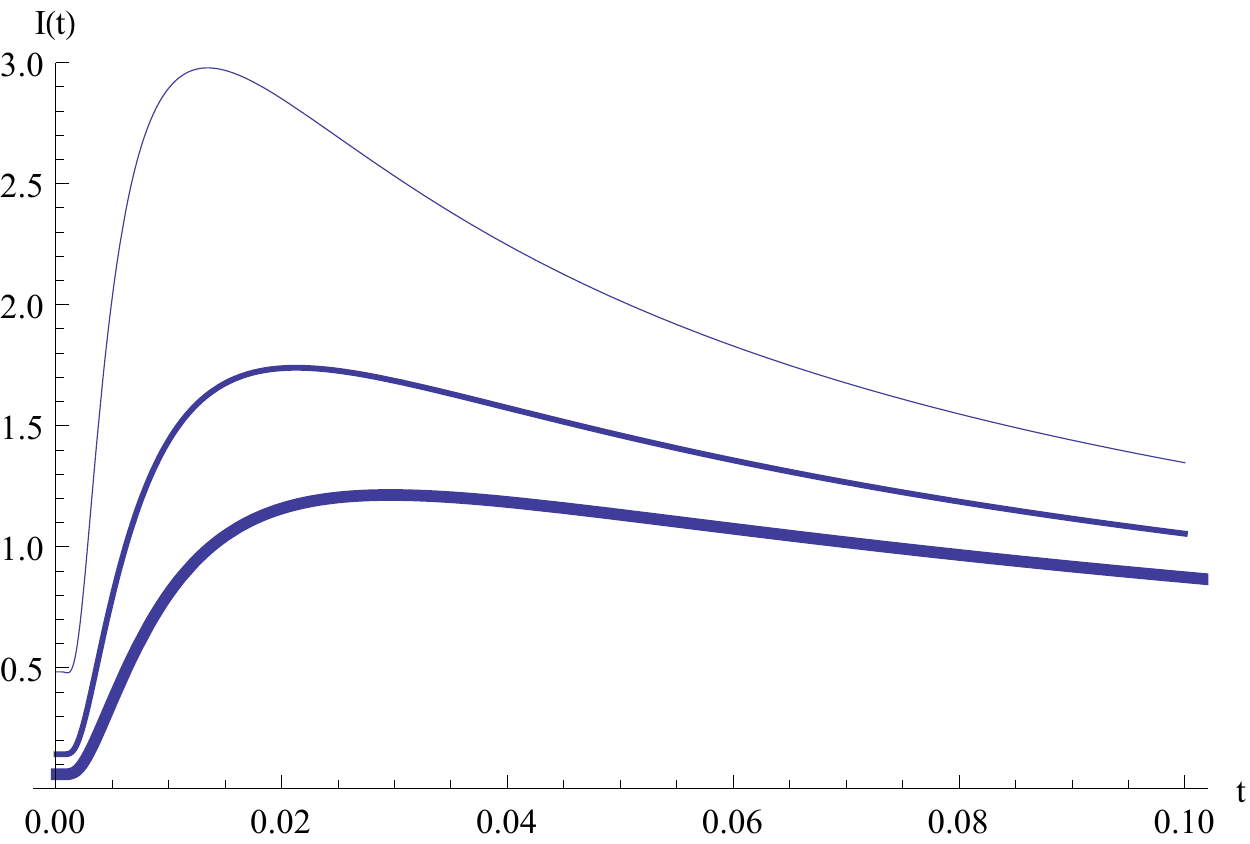}
\end{center}
\caption{The  dissipative impedance  for  {Majorana  particle-hole} contribution  $\frac{\zeta_{22,01}(\vec{Q},\Omega;0,\neq0)}{10^{-10}}\approx  (\frac{Q_{1}}{q})(\frac{Q_{2}}{q}) 10^{-2}I(t)\Big[\frac{Newton\cdot sec.}{m^3}\Big]$. The range of  temperature is $0.005<t<0.1$ (temperature $t=1$ corresponds to $0.1$ meV and the structure for  $t<0.005$ is an artifact of the numerics).}
\end{figure}

The  Majorana contribution  for  the particle-particle part [see Eq. $(22)$] for the same values of $\epsilon(b,k_{F})$   and tunneling amplitude   $t_{0}$ as in Fig. $(2)$ shows an absorption for low temperatures.
\begin{figure}
\begin{center}
\includegraphics[width=3.5 in ]{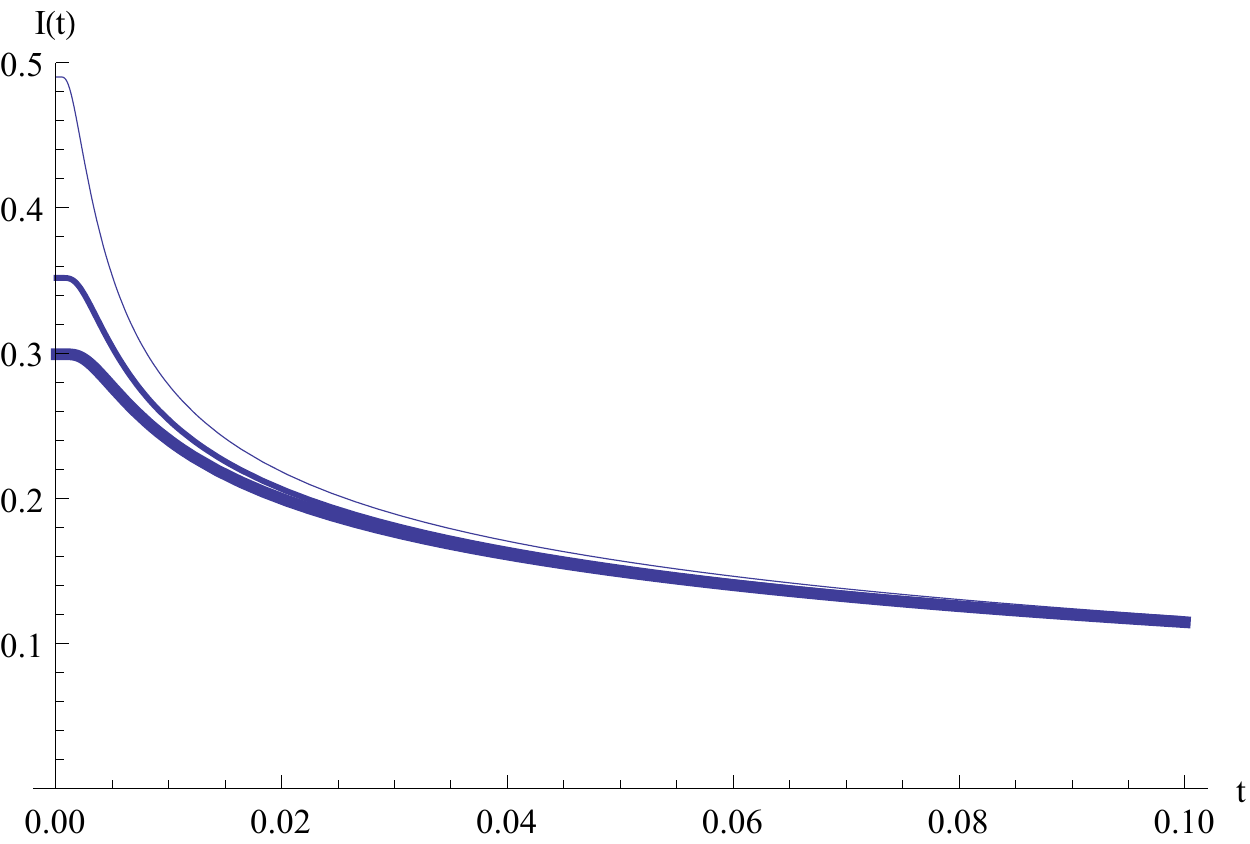}
\end{center}
\caption{The  dissipative impedance  for {Majorana  particle}  contribution  $\frac{\zeta_{22,22}(\vec{Q},\Omega;0,\neq0)}{10^{-10}}\approx (\frac{Q_{1}}{q})(\frac{Q_{2}}{q}) 10^{-1}I(t)\Big[\frac{Newton\cdot sec.}
{m^3}\Big]$   as a function of    $\epsilon(b,k_{F})= 0.005$ meV  (thin line), $\epsilon(b,k_{F})= 0.0075$ meV, $\epsilon(b,k_{F})= 0.01$ meV (thick line). The range of temperature is $0.005<t<0.1$ (temperature $t=1$ corresponds to $0.1$ mev and the structure for  $t<0.005$ is an artifact of the numerics).} 
\end{figure}
Comparing Fig. $(1)$ with Figs. (2), (3) we observe that the Majorana fermion gives rise to absorption at  low temperatures, in a region  where the particle-hole absorption (Fig. $1$) is absent.

Since impurities are always present, it is important to know    how to differentiate between the impurity  absorption  and the  Majorana fermions. Impurities will give rise to absorption for frequencies $\Omega>\epsilon(b,k_{F})-\epsilon_{impurity}$ and for temperatures $T>\Omega$; on the other hand, the Majorana absorption persists at  $T\rightarrow 0$ (see Fig. $3$).The total impedance is given by the sum of  contributions in the three figures (1) - (3).
We therefore conclude that the information about the Majorana modes, the magnetic field and tunneling amplitude $t_{0}$ can be obtained from the viscosity stress measurement.

\vspace{0.2 in}

\textbf{VII. Comments on  the  sound wave  analog of Andreev crossed reflection}

\vspace{0.2 in}

Next we comment on the   sound wave analog of the Andreev reflection.
 Two Majorana modes located  at the  two ends of a $p$-wave superconductor wire  are  detectable by  piezoelectric transducers   representing the sound equivalent of the   two-leads   experiments    which  measure  the    Andreev crossed reflection    \cite{Flensberg,DavidMajorana}.
We demonstrate that the same equations which were obtained for the Andreev crossed reflection    \cite{Flensberg,DavidMajorana} induced by a voltage between the two tips are obtained  for a sound wave  which  creates a time-dependent  lattice deformation  $D(t)=u(\frac{L}{2},t)+u(-\frac{L}{2},t)$. Here $u(\pm\frac{L}{2},t)$ is the sound deformation in the vicinity of each tip. The lattice deformation acts as a bias field. The voltage field $e^{\pm  i  \frac{ e}{\hbar} V t}$ in the two-tip experiment is replaced by  a bias field $e^{\pm ik_{F}D(t)}$ for the sound wave case.

We follow the derivation given in refs.  \cite{Flensberg,DavidMajorana}.
For a $p$-wave (or equivalently a one-dimensional wire with spin-orbit interaction in the proximity of an $s$-wave superconductor  and  a magnetic field) with length $L$ we have two Majorana modes localized at  $x=-\frac{L}{2}$ and   $x=\frac{L}{2}$. The  fermions for the two tips are represented by  $C_{1}(x=-\frac{L}{2})=e^{ik_{F}(-\frac{L}{2})}R_{1}(x=-\frac{L}{2})+e^{-ik_{F}(-\frac{L}{2})}L_{1}(x=-\frac{L}{2})$ and $C_{2}(x=\frac{L}{2})=e^{ik_{F}(\frac{L}{2})}R_{2}(x=\frac{L}{2})+e^{-ik_{F}(\frac{L}{2})}L_{2}(x=\frac{L}{2})$.  (Here $R_{1}$, $R_{2}$, $L_{1}$ and $L_{2}$ are the  right and left  chiral fermions and $k_{F}$ is the Fermi  momentum of the electrons).

Following ref. \cite{DavidMajorana} we integrate the Majorana fermions and obtain the coupling between the two tips:
$H_{eff.}(t)=(-ig^2)\int_{0}^{\infty}\,d\tau\chi^{+}(t)e^{-i\epsilon_{0}\tau}\chi(t-\tau)$ where $\epsilon_{0}$ is the overlap energy  between the two Majoranas.  Ignoring the oscillating  terms $ e^{ik_{F}(\pm\frac{L}{2})}$ allows us to simplify the form of $H_{eff.}$. In order to study the response to  sound waves  we replace $e^{\pm ik_{F}(\pm\frac{L}{2})}\rightarrow e^{\pm ik_{F}(\pm\frac{L}{2}+u(\pm\frac{L}{2}))}$,  where  $u(\pm\frac{L}{2})$  is the sound deformation induced  by the transducer.
 The deformation  field $D(t)=u(\frac{L}{2},t)+u(-\frac{L}{2},t)$ is a function of the  transducer frequency $\Omega$. The velocity strain field is given by $\epsilon_{01}(t)=\partial_{t}D(t)$.   The  derivative of  the  effective Hamiltonian $ H_{eff.}(t)$  with respect to the strain  velocity  $\epsilon_{01}(t)$  determines the    momentum density   $\pi(x,t)$,

$\pi(x,t)=\frac{\partial H_{eff.}(t)}{\partial (\partial_{t}D(t))}\approx (\frac{-ig^2 k_{F}}{\epsilon_{0}})\theta[t] \Big[ e^{ik_{F}D(t)}J(-\frac{L}{2},\frac{L}{2};t)+ e^{-ik_{F}D(t)}J^{*}(-\frac{L}{2},\frac{L}{2};t)\Big]$

\noindent where
$$J\left(-\frac{L}{2},\frac{L}{2};t\right)\equiv L^{\dagger}_{2}\left(\frac{L}{2},t\right)R_{1}\left(-\frac{L}{2},t\right)-L^{\dagger}_{1}\left(-\frac{L}{2},t\right)R_{2}\left(\frac{L}{2},t\right)+R_{1}\left(-\frac{L}{2},t\right)R_{2}\left(\frac{L}{2},t\right)$$
$$+L^{\dagger}_{2}\left(\frac{L}{2},t\right)L^{\dagger}_{1}\left(-\frac{L}{2},t\right).$$ 
 \noindent Here  $J(-\frac{L}{2},\frac{L}{2};t) $ is the correlation between the tips; when a  voltage is applied between the tips this correlation represents the current operator  
 \cite{DavidMajorana}.

From the  momentum density  we  compute the dissipative viscosity and the time ordered  correlation function,  $ R_{01,01}(q,\Omega)= (\frac{-i}{\hbar})\langle|T\Big(\pi(q,t)\pi(-q,0)\Big)|\rangle$. Following Eq. $(18)$ we obtain the viscosity  $ \zeta_{01,01}(q,\Omega)=-\frac{R_{01,01}(q,\Omega)}{i\Omega}$.
 The  viscosity  $\zeta_{01,01}(q,\Omega)$ is equivalent to the   Andreev crossed reflection conductance  obtained when   voltage is applied  between  the tips  \cite{Flensberg,DavidMajorana}.  
 [To compare the two correlation functions  we need to replace $\frac{e}{\hbar}Vt$ with $k_{F}D(t)$ and $g^2 e$ with $\frac{ g^2k_{F}}{\epsilon_{0}}$.]

\vspace{0.2 in}

\textbf{VIII. Conclusions}

\vspace{0.2 in}

\noindent 
In the first part of this paper we derived the spinor solution for an Abrikosov vortex lattice in a topological superconductor. We then obtained the zero and non-zero mode wave functions.
These  results have been used to compute the dissipative viscosity, which is obtained as a stress response to an applied velocity strain field. Experimentally one uses   two transducers, one for measuring the stress response and the second  transducer to  generate the strain field.
We  find in addition to the particle-hole contribution, a viscosity term which reflects the presence of Majorana fermions.
Probing the  $p$-wave wire with  a sound wave one thus finds  an effect similar to the Andreev crossed reflection.

\vspace{0.3 in}

\textbf{Apendix A. Vortex lattice }

\vspace{0.1 in}

\noindent
We consider the   Hamiltonian given by Eq. $(2)$ in a magnetic field.
\begin{eqnarray}
&&\hat{H}= \int\,d^2x \hat{\Psi}^{\dagger}(\vec{x})\Big[\tau_{3}\hat{h}_{3}+\tau_{2}\hat{h}_{2}+\tau_{1}\hat{h}_{1}\Big]\hat{\Psi}(\vec{x}),\hspace{0.025in}\frac{1}{2}\vec{\partial}_{x}\theta(\vec{x})\equiv\frac{1}{2}\sum_{l}\vec{\partial}_{x}\varphi_{l}(\vec{x}),
\varphi_{l}(\vec{x})\equiv \arg\Big(\vec{x}-\vec{R}_{l}\Big)\nonumber\\&&
\hat{h}_{3}=\frac{ v}{2k_{F}}\Big(\Big(-i\vec{\partial}_{x} -\frac{e}{\hbar} \vec{A}(\vec{x})+\frac{1}{2}\vec{\partial}_{x}\theta(\vec{x})\Big)^2- k^2_{F}\Big),\hspace{0.025in} \hat{h}_{2}= i\frac{|\Delta(\vec{x})|}{2k_{0}} \partial_{1},\hspace{0.025in}\hat{h}_{1}= -i\frac{|\Delta(\vec{x})|}{2k_{0}} \partial_{2} .\nonumber\\&&
\end{eqnarray}
 We consider first a single vortex  in Eq. $(2)$  and find from the London equation that the effective magnetic field $b$  is replaced by a string-like solution \cite{Ezawa,Tinkham},
$b(\vec{x})=\frac{\Phi_{L}}{2\pi \lambda^{2}_{L}}K_{0}\Big(\frac{|\vec{x}-\vec{R}_{l}|}{\lambda_{L}}\Big)$ which vanishes for  $|\vec{x}|\gg d$ ($d$ is the vortex lattice constant, $\lambda_{L}$ is the magnetic penetration depth and $K_0$ is the modified Bessel function). Since  we are interested in constructing a periodic   solution we replace the magnetic field  for  $|\vec{x}|<d\approx \lambda_{L}$  with a constant magnetic field. The constant magnetic field restricted to $|\vec{x}|<d$ is approximated by the spatial average   around the core with a radius of $d$.
\noindent
In order to study the vortex properties we  add to the  superconducting  Hamiltonian $\hat{H}$ [given in Eq. $(2)$]  the magnetic energy $ H_{M}$  and  the condensation energy  $H_{\Delta}$:
 \begin{eqnarray}
&&H_{eff.}=\hat{H}+H_{M}+H_{\Delta} ,\nonumber\\&&
H_{M}=\int\,d^2x \left(\frac{1}{2\mu_{M}}\right)\Big[\partial_{1}A_{2}(\vec{x},t)-\partial_{2}A_{1}(\vec{x},t)\Big]^{2} ,\nonumber\\&&
H_{\Delta}=\frac{1}{g}\int\,d^2x|\Delta|^{2} ,\nonumber\\&&
\end{eqnarray}
where $A(x,t)$ denotes the vector potential and $\Delta$ is the gap parameter.

\noindent

The variation of Hamiltonian $H_{eff.}$ in Eq.$(26)$  determines  the London equation for a single vortex.
\begin{eqnarray}
&&-\lambda^2_{L}b(\vec{x})+b(\vec{x})=\Phi_{L}\delta[\vec{x}-\vec{R}_{l}] , \nonumber\\&&
\lambda^2_{L}=\frac{k_{F}}{\mu_{M}e^2 \rho }, \hspace{0.1 in}    \rho= \frac{ v}{k_{F}} . \nonumber\\&&
\end{eqnarray}
The  solution of the London equation is: $b (\vec{x})=\frac{\Phi_{L}}{2\pi \lambda^{2}_{L}}K_{0}\Big(\frac{|\vec{x}-\vec{R}_{l}|}{\lambda_{L}}\Big)$  where $K_{0}$ is the modified Bessel function  \cite{Tinkham}.
This string-like solution gives rise to the effective vector potential $A^{(i)}_{eff.}=2e\Phi_{L}\epsilon_{i,j}\frac{{x}_{i}}{2\pi r^2}\Big(\frac{|\vec{x}|}{\lambda _{L}}K_{1}(\frac{|\vec{x}|}{\lambda_{L}}) -1\Big)$, $i=1,2$,  $j=1,2$ ($\epsilon_{i,i}=\epsilon_{i,i}=0$, $\epsilon_{1,2}=1$, $\epsilon_{2,1}=-1$).
As a result  the  field  $ |\hat{\Psi}_{\neq 0}(\vec{x})|^2$  vanishes  at $ \vec{x}=\vec{R}_{l}$  and is constant at  large distances.  
\noindent
For a {\it periodic structure of vortices}  we find
\begin{equation}
 -\lambda^2_{L}b(\vec{x})+b(\vec{x})=\Phi_{L}\sum_{l}\delta[\vec{x}-\vec{R}_{l}] . 
\label{periodic} 
\end{equation}
This equation  shows  that  the   effective magnetic field   $b(\vec{x})$ for multi-vortices  is periodic.

\vspace{0.1 in}

\textbf{Appendix B. Majorana fermion localized at the origin}

\vspace{0.1 in}

\noindent 
For a vortex which  is localized at $\vec{x}=0$, the phase  is $\varphi$.  The vortex field $\theta_{v}(\vec{x})=\theta_{v}(r,\varphi)$  obeys $[\theta_{v}(r,\varphi=2\pi)-\theta_{v}(r,\varphi=0)]=2\pi N$ with $N=1$  \cite{Gurarie}.
 The field  $\hat{C}(\vec{x})=\hat{C}(r,\varphi)$      transforms as   $\hat{C}(r,\varphi +2\pi)= -\hat{C}(r,\varphi)$ and $\hat{C}^{\dagger}(r,\varphi +2\pi)= -\hat{C}^{\dagger}(r,\varphi)$ in the presence of a multi-valued phase  $\theta(\vec{x})$.  Consequently, for  a $2\pi$ 
vortex, the fields $\hat{C}(\vec{x}) $, $\hat{C}^{\dagger}(\vec{x})$ are double-valued.  This corresponds to a half-vortex,  and a rotation of $2\pi$ implies a change of phase of $\pi$ in the fermion field.
Using polar coordinates, we have $(\partial_{1}+i\partial_{2})\rightarrow e^{i\varphi}(\partial_{r}+\frac{i}{|\vec{x}|}\partial_{\varphi})$ and $(\partial_{1}-i\partial_{2})\rightarrow e^{-i\phi}(\partial_{x}-\frac{i}{|\vec{x}|}\partial_{\varphi})$. The kinetic energy  in the presence of a vortex  at the origin $\vec{x}=0$  becomes,
\begin{equation}
K.E. \approx
\frac{v}{2k_{F}}\Big( -i\vec{\partial}_{x}\Big)^2-\mu_{eff.}(\vec{x}) . 
\label{equation}
\end{equation}
 At the origin 
  the effective chemical potential is positive far from  the vortex  and  changes sign for  $\vec{x}\rightarrow 0$. (A Majorana mode exists at the boundary of a trivial superconductor $\mu_{eff.}<0$ and a topological one with  $\mu_{eff.}>0$), 

\begin{eqnarray}
\left[\begin{array}{rrr}
-\mu_{eff.}(\vec{x}),&\frac {-|\Delta| }{2k_{0}}e^{i\varphi}(\partial_{x}+\frac{i}{r}\partial_{\varphi})\\
\frac {-|\Delta| }{2k_{0}}e^{-i\varphi}(\partial_{x}-\frac{i}{r}\partial_{\varphi}),&\mu_{eff.}(\vec{x})\\
\end{array}\right]
\left(\begin{array}{cc}U_{0}(r,\varphi)\\V_{0}(r,\varphi)
\end{array}\right)=0 . \nonumber\\&&
\end{eqnarray}
The zero mode solution is given by: $W_{0}(r,\varphi)=\Big[U_{0}(r,\varphi),V_{0}(r,\varphi)\Big]^{T}\equiv\Big[\frac{1}{\sqrt{i}}e^{\frac{i }{2}\varphi},\frac{1}{\sqrt{-i}}e^{\frac{-i }{2}\varphi}\Big]^{T}\frac{F(r)}{\sqrt{r}}$.  
The function $\frac{F(r)}{\sqrt{r}}$ obeys the normalization condition $\int\,d^2 r [\frac{F(r)}{\sqrt{r}}]^{2}<\infty$. 
For a vortex  lattice we use the Majorana solution which is given at $x=0$ and is translated periodically.

\vspace{0.1 in}


\noindent

\textbf{Appendix C.  Invariance of the Hamiltonian  under the coordinate transformation: derivation of  the stress-strain Hamiltonian}

\vspace{0.1 in}

\noindent
The viscosity tensor is obtained  from the  linear response of electrons in an   external field \cite{Fetter}. In order to accomplish this task we need to identify the elastic analog of the external electromagnetic field.  This is accomplished using the  invariance of the action under the coordinate transformation. 

The unperturbed crystal is described by the coordinates $\vec{x}$
and the deformed crystal by the coordinates  $\vec{\xi}$.
The distortion of the crystal is given by     $\vec{u}$  which is  caused either by  the phonons of the crystal  or by an  external force. We use the system  $\xi^{a}$  to describe the orthonormal coordinates for the deformed crystal with  the basis  vector frame $\partial_{\xi^{a}}$, a=1,2,3. The unperturbed crystal is described by the Cartesian coordinates $ x^{i}$ with  the basis   frame   vectors $\partial_{x^{i}}$, i=1,2,3.  The two coordinate  systems in the two frames   are related \cite{Katanaev}:
\begin{eqnarray}
&&\vec{x}\rightarrow \vec{\xi}(\vec{x})= \vec{x}+\vec{u}(\vec{\xi}) , \nonumber\\&& \xi^{a}(\vec{x})= x^{a}+u^{a}(\vec{\xi}), ~~a=1,2,3 ; \hspace{0.15 in} \xi^{i}(\vec{x})= x^{i}+u^{i}(\vec{\xi}), ~~i=1,2,3 , \nonumber\\&&
\end{eqnarray}
where $i$ represents the Cartesian  coordinates  and $a$ represents the deformed crystal. When the deformation of the crystal $\vec{u}$ vanishes  we have the relation $\vec{x}=\vec{\xi}$.
This allows us to introduce  the {non-relativistic}  transformation of the derivatives:
\begin{eqnarray}
&&\partial_{\xi^{a}}=\sum_{i=1,2,3}\frac{\partial x^{i}}{\partial_{\xi^{a}}}\partial_{i}=\sum_{i=1,2,3}\frac{\partial( \xi^{i} -u^{i})}{\partial_{\xi^{a}}}\partial_{i}=\sum_{i=1,2,3}
(\delta_{i,a}-\partial_{a}u^{i})\partial_{i}
;a=1,2,3 , \nonumber\\&& 
\frac{d}{dt}=\partial_{t}+\sum_{i=1,2,3}\partial_{t}u^{i}\partial_{i} .\nonumber\\&&
\end{eqnarray}
\noindent
 The metric integration for the deformed space  is   given in terms of the crystal deformation vector field   $\vec{u}$. That is, 
$dtd\xi^{a=1}d\xi^{b=2}d\xi^{c=3}= dt\mathbf{e}dx^{1}dx^{2}dx^{3} ,$ 
where $\mathbf{e}$ is the Jacobian of the coordinate transformation which  
for  the two-dimensional case  is given by: $\mathbf{e}=1-(\partial_{1}u^{1}+\partial_{2}u^{2})-\partial_{1}u^{1}\partial_{2}u^{2}$. We replace $\partial_{a}u^{j}\approx \partial_{i}u^{b} \delta_{a,i}\delta_{j,b} \approx \partial_{i}u^{j}$ (the exact relation is given by the matrix equation 
$\sum_{j}\partial_{a}u^{j}\partial_{j}u^{b}=\delta_{a,b}$).
 We   introduce the notation $\epsilon_{i,j}=\frac{1}{2}\Big(\partial_{i}u^{j}+\partial_{j}u^{i}\Big)$, $\epsilon_{0,i}=\partial_{t}u^{i}$ and $\Omega_{i,j}=\frac{1}{2}\Big(\partial_{i}u^{j}-\partial_{j}u^{i}\Big)$. When  the excitations are caused by the phonons, we use the phonon spectrum of the crystal (in the harmonic representation).  Due to the compatibility conditions \cite{Kosevich}  we have  $\Omega_{i,j}=0$ and for a crystal with disclinations we have $\Omega_{i,j}\neq0$.

\end{document}